\documentclass[pre,aps,twocolumn,showpacs,floatfix,10pt,superscriptaddress]{revtex4-1}
\usepackage{graphicx}
\usepackage{color}
\usepackage{amsmath, amsthm, amssymb}
\usepackage{epsfig}
\usepackage[utf8]{inputenc}
\usepackage[T1]{fontenc}
\newcommand{\be}{\begin{equation}}
\newcommand{\ee}{\end{equation}}
\newcommand{\bea}{\begin{eqnarray}}
\newcommand{\eea}{\end{eqnarray}}
\newcommand{\beq}{\begin{equation}}
\newcommand{\eeq}{\end{equation}}
\newcommand{\beqa}{\begin{eqnarray}}
\newcommand{\eeqa}{\end{eqnarray}}
\newcommand{\oa}{\overset{A}\vert}
\newcommand{\ob}{\overset{B}\vert}

\begin{document}

\title{Logarithmic speed-up of relaxation in A-B annihilation with exclusion}
\author{Rahul Dandekar}
\email{dandekar@theory.tifr.res.in}
\affiliation{The Institute of Mathematical Sciences - HBNI, CIT Campus, Taramani, Chennai, India}


\begin{abstract}
	We show that the decay of the density of active particles in the reaction $A+B \rightarrow 0$ in one dimension, with exclusion interaction, results in logarithmic corrections to the expected power law decay, when the starting initial condition (i.c.) is periodic. It is well-known that the late-time density of surviving particles goes as $t^{-1/4}$ with random initial conditions, and as $t^{-1/2}$ with alternating initial conditions ($ABABAB$...). We show that the decay for periodic i.c.s made of longer blocks ($A^{n}B^{n}A^{n}B^{n}$...) do not show a pure power-law decay when $n$ is even. By means of first-passage Monte Carlo simulations, and a mapping to a q-state coarsening model which can be solved in the Independent Interval Approximation (IIA), we show that the late-time decay of the density of surviving particles goes as $t^{-1/2}(\log{(t)})^{-1}$ for $n$ even, but as $t^{-1/2}$ when $n$ is odd. We relate this kinetic symmetry breaking in the Glauber Ising model. We also see a very slow crossover from a $t^{-1/2}(\log{(t)})^{-1}$ regime to eventual $t^{-1/2}$ behaviour for i.c.s made of mixtures of odd- and even-length blocks.
\end{abstract}


\maketitle

\section{Introduction} The two-species annihilation reaction $A+B \rightarrow \phi$ is one of the simplest examples of a reaction-diffusion system. It has been studied for over thirty years, after first being introduced by Zeldovich and Ovchinnikov in the context of bimolecular chemical reactions \cite{zeldovich78}, and by Toussaint and Wilczek \cite{toussaintwilzek} in the context of particle-antiparticle annihilation in the early universe, and has also been applied to hole-particle annihilation in irradiated semiconductors\cite{argkop90} . The reaction has also been important in clarifying the breakdown of mean-field kinetics for finite dimensions \cite{avrahamredner,kangredner}. In particular, it was realised that in dimensions less than $4$, the long-time decay starting from an initially well-mixed configuration does not conform to the mean-field kinetics.

Consider the process $A+B \rightarrow \phi$, starting from initial conditions created by an independent Bernoulli point process on each site of d-dimensional lattice: each site holds either an A-type particle, a B-type particle, or is empty, with probabilities $c_A$, $c_B$, and $(1-c_A-c_B)$ respectively. We concentrate on the case where the $c_A=c_B$. Bramson and Lebowitz proved that for large times $t$, for $d<4$, the density of surviving particles decays as $t^{-d/4}$. This result differs from the mean-field result which predicts a decay proportional to $t^{-1}$. The mean-field result fails because it assumes well-mixed concentrations of both reactants, whereas the true picture, as elaborated on later by Leyvraz and Redner \cite{leyvrazredner}, is that of a diffusion-controlled late-time separation of the system into domains of A-rich and B-rich regions. Due to the random nature of the initial condition, a large region of size $L$ has an excess of one type of particle over the other, of magnitude $O(L^{1/2})$. Since the average length of the domains increases diffusively as $L \sim t^{1/2}$, this gives a decaying density of $t^{-1/4}$.

This domain picture fails for correlated initial conditions where the excess of one particle type over the other in a large region is $O(1)$, for example, period i.c.s like AABBAABBAABB... . In this case, we use another picture of the late-time process. At late times, if it is reasonable to expect that A and B particles mix well enough such that they encounter each other equally often, we can remove the distinction between A and B type particles, accompanied by a halving of the reaction rate. The long time decay with equal starting densities would be then be given by the kinetics of the single-species annihilation reaction \cite{singlespec}, and hence goes as $t^{-d/2}$ for $d<2$. 

In this paper, we show that even this mixing requirement fails in case of particles with exclusion interaction in 1D. We show that, surprisingly, the decay from intial conditions of the kind $A^{n}B^nA^nB^n$ depends on the parity of $n$. For odd $n$ there is an eventual pure $t^{-1/2}$ decay, whereas for odd $n$ the density of particles decays as $t^{-1/2}/\log{(t)}$ in the long-time limit. 

A variety of exact results available for simple models such as the symmetric exclusion process\cite{derridarevs} have demonstrated the drastic effects of exclusion interaction between particles in one dimension. For example, a tagged particles in the simple exclusion process in 1D shows subdiffusive behaviour \cite{taggedparticle1}\cite{taggedparticle4}\cite{taggedparticle5}.

Deviations from the expected behaviour because of exclusion have been observed in simulations for various systems which map to the AB annihilation problem with exclusion. Ódor and Menyhárd \cite{odor} studied systems of interacting particles for initial conditions which map to the initial conditions $A^{n}B^{n}A^{n}B^{n}...$ in the AB annihilation problem, where $[A^{n}]$ denotes a contiguous block of $n$ A-type particles. They studied the behaviour for initial conditions which map to the periodic i.c.s $A^{n}B^{n}\dots$, and observed a decay exponent which is different from the expected value of $-0.5$. They also studied i.c.s which were made of random mixtures of blocks $A^{n}B^{n}$ with different lengths $n$, and in this case the observed decay exponents was found to vary with the proportions of different blocks in the mixture. They also argue that the exclusion interaction leads to a marginal (in the field theoretical sense) interaction between the particles, which could lead to a logarithmic correction to the decay. Recent simulations by Lee of the Conserved Lattice Gas (CLG), the dynamics of which also maps to AB annihilation with exclusion, showed an anomalous decay exponent of $0.523$ was observed for i.c.s which map to periodic i.c.s of the AB process with even-length blocks \cite{leeclg14}. 

These observed anomalous decay exponents for this important problem have not received an analytical explanation till now.  We show in this paper, that for the 1D AB annihilation problem with exclusion, and using the Independent Interval Approximation (IIA), that for a purely periodic initial condition made of blocks $A^{n}B^{n}$, the particle density in the long-time limit decays for even $n$ goes as $t^{-1/2} (\log{(t)})^{-1}$, while for odd $n$ one obtains the usual $t^{-1/2}$ decay. We also show that for i.c.s which are made of mixtures of blocks of different length, there is a long-time crossover between the logarithmic correction and the pure $t^{-1/2}$ decay, which can account for the varying exponent seen by Ódor and Menyhárd, and the exponent seen by Lee for natural initial conditions. Since the 1D CLG is an important simple model of an active-absorbing phase transition \cite{jain}\cite{kkclgtransition}, we discuss it in more detail in the conclusions.

We trace the origin of the logarithmic speed-up of the decay to the phenomenon of `kinetic symmetry breaking', studied in the Glauber Ising model by Majumdar, Dean and Grassberger \cite{majumdardean}. In section II, we briefly  review this study with a view to setting up a similar treatment of the AB system in the rest of the paper. The rest of the paper is devoted to studying the AB annihilation process starting from periodic i.c.s. In section III we consider block i.c.s $A^nB^nA^nB^n\dots$ with $n=1, 2, and 3$. We generalise to longer blocks in section IV. In section V we briefly discuss the results of First-Passage Monte-Carlo simulations of the model, and in section VI we generalise our reasoning to i.c.s made of mixtures of blocks of different lengths, showing that under general conditions, there is a long-time crossover to a pure $t^{-1/2}$ decay. We conclude in section VII along with a discussion of how our results apply to recent simulations on the 1D Conserved Lattice Gas by various authors.

\section{Kinetic Symmetry Breaking in the Glauber model} 

Our aim in the paper is to study the two-species annihilation reaction with exclusion. Each site can only be in three states - occupied by an A particle, occupied by a B particle, or empty (denoted by $\phi$). The particles react and diffuse according to the rules
\beqa
AB &\overset{1}\rightarrow \phi\phi \label{eq:rule1}\\
A\phi &\overset{1/2}\rightleftharpoons \phi A \\
B\phi &\overset{1/2}\rightleftharpoons \phi B \label{eq:rule3}
\eeqa

We study the AB annihilation reaction through a nontrivial mapping to Glauber Ising dynamics with kinetic symmetry breaking \cite{majumdardean}. In this section we briefly review the results of \cite{majumdardean}, and also set up notation for the rest of the paper. Consider the 1D Ising model, and denote by $W(S_{i-1},S_i,S_{i+1})$ the rate of flip of the $i^{th}$ spin, $S_i \rightarrow -S_i$ when the neighbouring spins are $S_{i-1}$ and $S_{i+1}$. Majumdar, Dean and Grassberger studied the case with the following modified zero-temperature Glauber rates:
\beqa
W(+,+,+)&=&W(-,-,-)=0\\
W(+,+,-)&=&W(+,-,-)=1/2\\
W(-,-,+)&=&W(-,+,+)=1/2\\
W(+,-,+)&=&\alpha\\
W(-,+,-)&=&1
\eeqa
For $\alpha=1$ one recovers up-down symmetry in the rates (and the zero-temperature Glauber model). For $\alpha<1$ the symmetry between up and down spins is broken. Denote the magnetisation per unit length by $m$ and the total magnetisation of the system by $M$. For all $\alpha$, the only stable states are those with $m=\pm 1$. We now argue that for all $\alpha<1$,  $m(t)\rightarrow -1$ as $t\rightarrow \infty$ for almost all initial conditions, in the thermodynamic limit.

For un-modified Glauber dynamics ($\alpha=1$), the total magnetisation $M$ performs a random walk on an axis from $M=-L$ to $M=+L$, ending at either value with equal probability. For $\alpha<1$, the rightward moves corresponding to annihilation of `$-$' (towards $M=+L$) occur with a rate $\alpha$, giving rise to a net leftward bias $r_1 \alpha L$ where $r_1$ is the density of `$-$' domains of unit length. Thus the system performs a biased random walk, that ends up at $M=-L$ with probability $1- O(e^-c L)$ for almost all initial conditions (all but those with $M=L-O(1)$).

We allow for the fact that each `$-$' domain might have an individual value of $\alpha$, a measure of how easy it is to annihilate that `$-$' domain (this value can be reassigned in some way when two domains merge). We denote by $\overline{\alpha}$ the value of $\alpha$ averaged over all the `$-$' domains in the system. Due to the merging of domains, $\overline{\alpha}$ can change with time.

We now outline the derivation in \cite{majumdardean}. By `number density' of a quantity, we mean the sum of the quantity over the lattice, divided by the lattice size. Denote the number density of domain walls by $N(t)$, and let $L_{+}(t)$ and $L_{-}(t)$ denote the number density of $+$ and $-$ domains respectively. Also, the number density of `$+$' domains of length $n$ will be denoted by $P_n$, and the corresponding for `$-$' domains by $R_n$. The dynamics obeys the exact equations $\frac{dN}{dt}=-P_1-\overline{\alpha}R_1$ and $\frac{dL_{+}}{dt}=-(1-\overline{\alpha})R_1$.

The Independent Interval Approximation (IIA) consists of replacing joint probabilities such as the probability to find a domain of length length $i$ next to a domain of length $j$, $P_{i,j}$, by the product of the probabilities of finding the two types of domains independently, $P_i P_j$ \cite{krapiviia}. The evolution equations for $P_n$ and $R_n$, in the IIA framework, are simpler to write in terms of the quantities $p_n = P_n/N$ and $r_n = R_n/N$, for $n\ge 2$ (for details see \cite{majumdardean}).

The evolution equations for $p_n$ and $r_n$ can be simplified for large times, when the system has a magnetisation per site close to $-1$. Consider the system at a late-time stage, when it is made of long `$-$' domains punctuated by small `$+$' domains. These `$+$' domains rarely merge, and hence the evolution of $p_n$ is dominated by diffusion of the domain walls. On the other hand, the dominant mechanism for the change in their length distribution of the `$-$'-domains turns out to be the merging (ie, annihilation of the intervening `$+$'-domains), which changes $r_n$ faster than the diffusion terms. Hence, at late times, keeping only the dominant terms, we have, for $n \ge 2$
\beqa
\frac{dp_n}{dt}&\approx& p_{n+1}+p_{n-1}-2p_n \label{eq:pnapprox}\\
\frac{dr_n}{dt}&\approx& p_1 \left(\sum_{i=1}^{n-2} r_i r_{n-i-1}-r_{n+1} \right) \label{eq:rnapprox}
\eeqa
with $\frac{dr_1}{dt} \approx r_2 - r_1$.

The first equation implies $\sum n p_n \approx c_1 \sqrt{t} = L_{+}(t)/N(t) \equiv l_{+}$. The second equation is solved by the ansatz $r_n = \lambda(t) \exp{(-n \lambda(t))}$ where $\lambda(t)$ obeys the equation $N(t)\frac{d\lambda}{dt} = \lambda(t)\frac{dN}{dt}$, which implies that $\lambda(t) = c_2 N(t)$. $c_1,c_2$ are constants set by the state of the system at a time $t_0$ when the IIA description starts to hold \cite{majumdardean}. 

Solving for $N(t)$ and $m(t)$ using $L_{+}(t) \approx c_1 N(t) \sqrt{t}$ and $r_1 \approx c_2 N(t)$, one gets,
\beqa
N(t) &=& \frac{c}{\sqrt{t} \log{(bt)}} \mbox{~and}\\
m(t) &=& -1 + \frac{a}{\log{(bt)}}
\eeqa
where $b \equiv t_0^{-1}$, and $a,c$ are constants which depend on the state of the system, particularly the values of $m(t_0)$ and $r_1(t_0)$ at a time $t_0$ sufficiently large that (\ref{eq:pnapprox}) and (\ref{eq:rnapprox}) hold. 

The above treatment holds even with the following two modifications, important in the next few sections: (a) if $\overline{\alpha}$ is a function of time, that approaches a non-zero constant value for large times $\overline{\alpha} \rightarrow c>0$, and (b) if the $+$ domains also have $\alpha_+< 1$, as long as $\alpha_-<\alpha_+$.

\section{Block initial conditions with blocks of lengths 1, 2 and 3} We now study the A-B annihilation process defined by eqns (\ref{eq:rule1})-(\ref{eq:rule3}). The dynamics follows continuous-time updates, and time is measured in the number of Monte-Carlo sweeps of the system. In this and the next section, we will study the decay of the number of surviving particles, starting from periodic initial conditions of the form A$^{n}$B$^{n}$A$^{n}$B$^{n}$..., where we call $n$ the block-length. In this section, we consider $n=1,2$ and $3$.

The decay of the CLG starting from an initial condition which corresponds to the i.c. ABABAB... ($n=1$), was first studied by Bandyopadhyay \cite{bandyopadhyay}. Kwon and Kim \cite{kwonkimclg} pointed out a simple mapping to the zero-temperature Glauber Ising model, by identifying the domain walls with particle types: $+|- \rightarrow A$ and $-|+ \rightarrow B$. Then the block-length 1 initial condition stated above maps to the starting configuration

\beq
+\oa - \ob + \oa - \ob + \oa - \ob + \oa - \ob + ... \label{eq:1bic}
\eeq 

Then the rules (\ref{eq:rule1} - \ref{eq:rule3}) map to the zero-temperature Glauber Ising evolution rules (without kinetic symmetry breaking, that is, with $\alpha=1$). The evolution of the density of domain walls in the system is known to be \cite{krapivskybook}
\beq
N(t) = \frac{1}{\sqrt{4 \pi t}} + O(\frac{1}{t})
\eeq

Now we consider the i.c. with $n=2$, AABBAABB... . We make a mapping of this system to a system where each dual-lattice site can take three values of `spin', $q=1,2,3$. Identify 
\beqa
&1&|2 \rightarrow A,~ 2|3 \rightarrow A, \mbox{~and~}\\
&2&|1 \rightarrow B,~ 3|2 \rightarrow B
\eeqa
An A-type domain wall increases the value of $q$, while a B-type domain wall decreases it. The $n=2$ i.c. becomes
\beq
1\oa 2 \oa3 \ob 2 \ob 1 \oa 2 \oa 3 \ob 2 \ob 1... \label{eq:2bic}
\eeq
\noindent where a vertical line denotes a domain wall. A possible state after one A-B annihilation step would be
\beq
1\oa 2\: \:  2 \: \: 2 \ob 1 \oa 2 \oa 3 \ob 2 \ob 1...
\eeq
The A-B annihilation dynamics in eqns. (\ref{eq:rule1}) - (\ref{eq:rule3}) thus maps a dynamics to the following dynamics for domain walls:
\beqa
q_1|q_2|q_1 &\overset{1}\rightarrow& q_1~ q_1~ q_1 \label{eq:evol1} \mbox{~and}\\
q_1~q_1|q_2 &\overset{1/2}\rightleftharpoons& q_1| q_2~q_2 \label{eq:evol2}
\eeqa

This is different from the traditional Glauber dynamics for the q-state Potts model \cite{pottsderrida}, in which domain walls can coalesce as well (coalescence is a process where two domain walls coalesce into one, of the kind $q_1|q_2|q_3 \rightarrow q_2~ q_2 |q_3$). In our mapping domain walls cannot coalesce, because of the exclusion between particles of the same species.

A domain with a given value of $q=i$, say, is called an $i$-domain. A 2-domain can be annihilated only if it has 1-domains on both sides, or 3-domains on both sides. Starting from the i.c. eqn. (\ref{eq:2bic}), none of the 2-domains can be annihilated at the first time-step due to exclusion. Thinking back to the kinetic symmetry-broken Ising model, we can thus assign a local value of $\alpha=0$ to a 2-domain lying between a 1- and a 3-domain, and $\alpha=1$ to one with same domains on either side. Starting from eqn. (\ref{eq:2bic}), one sees that 1-domains and 3-domains will always be surrounded by 2-domains on both sides, and hence have $\alpha=1$. Domain walls of the form $3|1$ and $1|3$ are never formed.

Now we can map the evolution of domains walls in this system to that of a corresponding Ising configuration with kinetic symmetry breaking, by labeling 1- and 3- domains as `$+$' domains and 2-domains as `$-$' domains, now allowed to have a local values of $\alpha$. The value of $\alpha$ for the 2-domains changes on merging according to the rule $\alpha^{new} = (\alpha^{(1)}+\alpha^{(2)})\mod{(2)}$.

We now argue that the value of $\overline{\alpha}$, which starts at $0$, converges at long times to $1/2$, never becoming $1$. At large times, the system is dominated by large 2-domains, punctuated by small 1- and 3-domains which rarely merge. The long-time value of $\overline{\alpha}$ is given by the steady-state value of $\alpha$ of the merging process $\alpha_1+\alpha_2 \rightarrow (\alpha_1+\alpha_2)\mod{(2)}$. In this steady-state, $\alpha=0$ and $\alpha=1$ are equally likely, giving $\overline{\alpha}_{t\rightarrow\infty}=1/2$.

Since this $\overline{\alpha}<1$, the results from the previous section apply, and we expect that
\beqa
N(t) \approx \frac{c}{\sqrt{t} \log{(bt)}}
\eeqa
The decay in the $n=2$ case thus has a logarithmic speed-up compared to the decay in the $n=1$ case.

Now consider the i.c. with $n=3$. Mapping to a 4-state model, the domain walls map to particle type as $q|(q+1) \rightarrow A$ and $q|(q-1) \rightarrow B$, and hence the i.c. $AAABBBAAABBB\dots$ becomes
\beq
1\oa 2\oa 3\oa 4\ob 3\ob 2\ob 1\oa 2\oa 3\oa 4\ob 3\ob 2\ob 1\dots \label{eq:3bic}
\eeq
The evolution rules are as in eqns. (\ref{eq:evol1}) and (\ref{eq:evol2}), with annihilation but no coalescence of domain walls.

We note a symmetry between the evolution of domains with $q=i$ and $q=5-i$, that is, $q$-values which are at the same `depth' inside A- or B-blocks in the eqn (\ref{eq:3bic}). The values of the annihilation rate $\alpha$ for various types of domains are calculated as folows: The 1- and 4- domains can always be annihilated if they reach unit length, giving $\overline{\alpha}_1 = \overline{\alpha}_4 = 1$ whereas for the 2- and 3- domains $\overline{\alpha}_2 = \overline{\alpha}_3 = \overline{\alpha}<1$ as $t\rightarrow \infty$.

Forgetting for the moment the distinction between 2- and 3- domains, we assign them both the label `$-$', and assign 1- and 4-domains the label `$+$'. After this labeling the i.c. becomes 
\beq
+|--|+|--|+|--|+|...
\eeq
where only domain walls of types $1|2$, $2|1$, $3|4$, and $4|3$ are visible. Call $N_1(t)$ the density of these domain walls. Then, for large times, $N_1(t)$ is upper-bounded by the kinetic-symmetry-broken Glauber decay
\beq
N_1(t) \le O(t^{-1/2}(\log{(t)})^{-1})
\eeq
Call $N_2(t)$ the density of domain walls of the types $2|3$ and $3|2$. The 2- and 3- domains coarsen within the large `$-$' domains in the labeling above. The average size of a `$-$' domain grows with time as $t^{1/2}\log{(t)}$. The 2- and 3- domains have the same value of $\alpha$, and thus there is no kinetic symmetry-breaking between them, and the average sizes of 2- and 3-domains thus grow diffusively, as $t^{1/2}$. (Therefore, each '$-$' domain of size $\sim t^{1/2} (\log{(t)})$ contains a large number of 2- and 3-domains, and this picture is consistent.) We then get $N_2(t) \approx c/\sqrt{t}$. The total number of domain walls decays as
\beq
N(t) = N_1(t)+N_2(t) \sim t^{-1/2} (1+O((\log{(t)})^{-1}))
\eeq

\section{Generalization to longer blocks} We now proceed to longer blocks ($n>3$), for which there is no simple relabeling scheme which allows one to directly use results from the Kinetic symmetry-broken Ising model. However, we use the insights of the preceding section to determine the form of the final uniform state, more precisely, whether it is always a particular q-value or if it is two q-values with equal probability, and then construct a mapping to the kinetic symmetry-broken Ising dynamics that is exact as $t\rightarrow \infty$. This allows us to determine the dominant term in the long-time decay.

Let us consider the $n=4$ i.c. In the q-state picture,
\beq
1\oa 2\oa 3\oa 4\oa 5\ob 4\ob 3\ob 2\ob 1\oa 2\oa 3\oa 4\oa 5\ob 4\ob 3\ob 2\ob 1\dots
\label{eq:4bic}
\eeq

We note that: (1) There is a symmetry between the evolution for $q$ and $(6-q)$-type domains, and (2) the 1- and 5- type domains will get annihilated quickly, the average distance between them growing at least as fast as $t^{1/2}\log{(t)}$.

We proceed to the determination of the asymptotic behaviour of $N(t)$ by showing three things:

(1) The final state is almost always the uniform state with $q=3$,

(2) The annihilation reaction rate for the 3-domains, $\overline{\alpha}_3$ approaches the value $1/2$ as $t\rightarrow \infty$, and

(3) The number of 2- and 4-domains is of the same order as the number of 5- and 1-domains, upper bounded by $c/(\sqrt{t}\log{(bt)})$.

Let us start with (1). We argue by contradiction. The final state cannot be $q=1$ and $q=5$, and if it is $q=4$, by symmetry, it should be $q=2$ with an equal probability, and hence the system at large times should be in a state with coarsening 4-type and 2-type domains of typical size $\sim t^{1/2}$ (this is the Leyvraz and Redner picture again). However, 4-type and 2-type domains cannot be in contact due to the un-eliminable 3-domains between them. However, the 2- and 4-domains {\it can} be eliminated by these 3-domains surrounding them on both sides, and hence the proposed late-time configurations are unstable to takeover by the 3-domains. Hence the final state is a uniform $q=3$ state with probability $1$ in the thermodynamic limit.

The late-time state looks (schematically) like this: 
\beq
33\dots333\ob222\oa33\dots33\oa4444\oa55\ob44\dots4\ob33\dots
\eeq
There are large 3-type domains, tiny and rare 1- and 5-domains, and 2- and 4-domains of a characteristic size as yet undetermined. The density of 1- and 5-type domains, call it $n_1(t)$, decays at least as fast as $1/(t^{1/2}\log{(t)})$. Denote the density of 2- and 4-type domains as $n_2(t)$ and that of 3-type domains as $n_3(t)$.

To show (2), namely that the value of $\overline{\alpha}_3$ approaches $1/2$ for large times, we can apply the same reasoning as in section III, namely that it is given by the steady-state value of the process $(\alpha^{old}_1 + \alpha^{old}_2)\mod{(2)} \rightarrow \alpha^{new}$.

We proceed to (3). We have two types of 2-domains: those with $\alpha=0$ have a 3-domain on one side and a 1-domain on the other, while those with $\alpha=1$ have 3-domains on both sides. (We can neglect the number of 2-domains with 1- or 5-domains on both sides, as that would require annihilation of 3-domains in between.) The same goes for 4-domains, with the result that $\overline{\alpha}_2 = \overline{\alpha}_4 \leq 1$ as $t \rightarrow \infty$. Assuming the Independent Interval Approximation, $\overline{\alpha}_2 = 1- n_1(t)/n_2(t)$. Now consider two cases:

(i) $\overline{\alpha_2} \rightarrow 1$ as $t \rightarrow \infty$. This means that $n_1/n_2 \rightarrow 0$, and hence 1- and 5-domains can be neglected. Then we have a system with domains of types 2, 3 and 4, where $\overline{\alpha_2} = \overline{\alpha_4} = 1$ and $\overline{\alpha_3} = 1/2$. This is the same as the eventual state for the i.c. $AABBAABB\dots$, with the relabelling of q-values $1 \rightarrow 2$, $2 \rightarrow 3$ and $3\rightarrow 4$.

(2) In the long time limit, $\overline{\alpha}_2$ is strictly less than 1. Since $\overline{\alpha}_2 = 1- n_1(t)/n_2(t)$, we get that $n_2(t) = \frac{1}{1-\overline{\alpha_2}} n_1(t) = \frac{c}{\sqrt{t}\log{(bt)}}$. 

Hence, in both cases, the total number of domains walls, $N(t) = 4 n_1(t) + 2 n_2(t) = \frac{c'}{\sqrt{t}\log{(b't)}}$.

A similar procedure can be followed for $n \geq 5$ i.c.s, to derive that

(a) for even $n$, there is a definite final state in the thermodynamic limit, the one with $q = n/2 + 1$. As a result, at large enough times domains with this value of $q$ have $\alpha=1/2$, while the density of other types of domains $\sim t^{-1/2} (\log{(t)})^{-1}$. Hence $N(t) \sim  \frac{C}{\sqrt{t} \log{(Bt)}}$.

(b) for odd $n$, the final state can be the uniform states $q = \lfloor n/2+1 \rfloor$ or $q= \lceil n/2+1 \rceil$ with equal probability. Hence $N(t) \sim t^{-1/2}$ for large $t$.

\begin{figure}[h]
\centering
\includegraphics[width=0.95\linewidth]{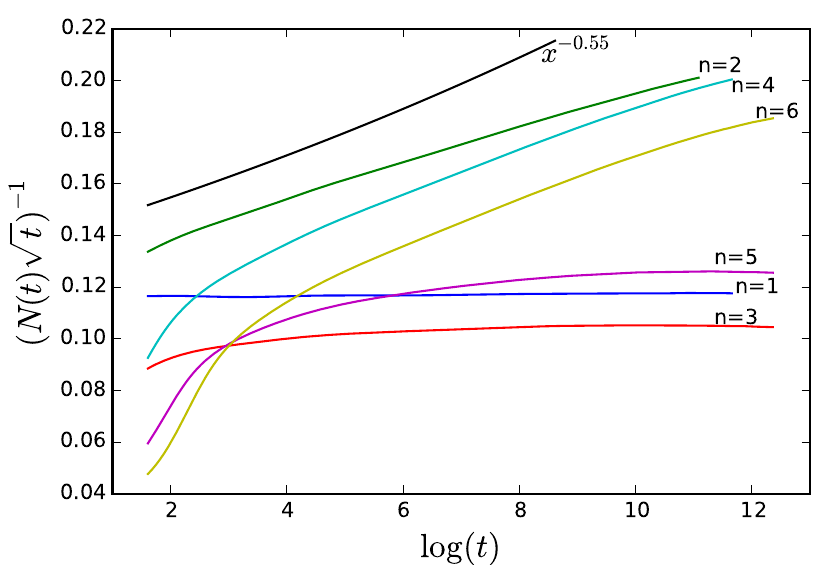}
\caption{A plot of $1/(N(t)\sqrt{t})$ (arbitrary units) against $\log{(t)}$, which either asymptotes to a straight line with a positive slope (even $n$), or to a constant (odd $n$). A pure power law decay is plotted for comparison - note the upward curvature of the line. For a lattice size $L=2$ x $10^7$, averaged over 500 realizations.}
\label{decay}
\end{figure}

\begin{figure}[h]
\centering
\includegraphics[width=0.95\linewidth]{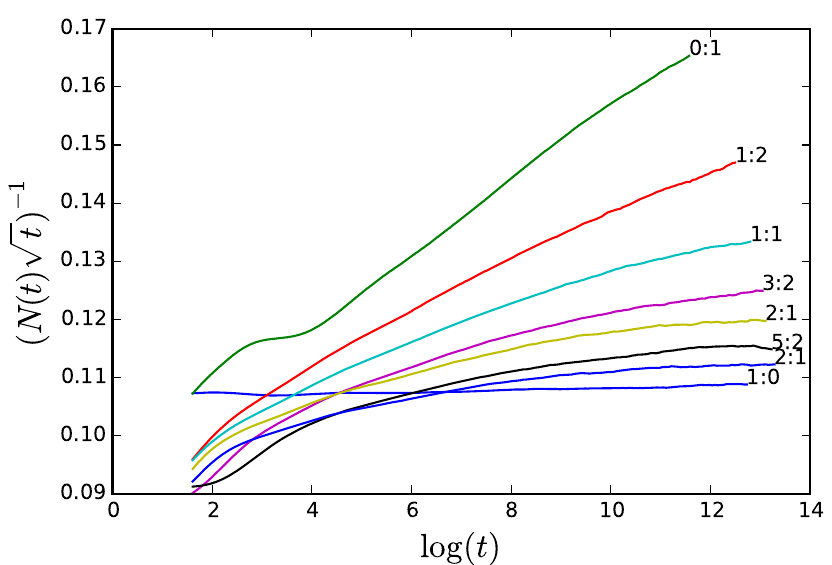}
\caption{A plot of $1/(N(t)\sqrt{t})$ (arbitrary units) against $\log{(t)}$, for initial conditions made of the units $(AB)^m(AABB)^n$, denoted by {m:n}. The results exhibit a crossover between intermediate-time $t^{-1/2}/\log{(t)}$ behaviour and eventual $t^{-1/2}$ behaviour, with crossover time depending on the ratio $n/m$. For a lattice size $L=10^7$, averaged over 100 realizations.}
\label{crossover}
\end{figure}

\section{First-passage Monte Carlo Simulations} We used a first-passage Monte Carlo algorithm \cite{kmc1} to run fast simulations of the decay starting from various initial conditions. Due to the dynamic nature of the algorithm, we could go to very large times, and use large lattice sizes. We used the lattice variant of the method, described in \cite{kmc2}.

Figure \ref{decay} shows the results of these simulations, starting from initial conditions with block lengths $n=1$ to $n=5$ on systems with $L=2$ x $10^7$, and periodic boundary conditions. We plot $(N(t)\sqrt(t)^{-1}$ versus $\log{(t)}$, showing tat it either asymptotes to a constant, or to a straight line on the logarithmic scale. We also studied the long-time behaviour (data not shown) of $\overline{\alpha}$ for various q-domains for $n=2,3,4$, and and the results are consistent with the discussion in the previous sections.

\section{ Mixed initial conditions} Define a `motif' as a finite-length string made of A's, B's and empty sites such that the number of A's is equal to the number of B's. Examples of motifs are the elementary blocks we have considered so far, viz. `AB', `AABB', etc. In this section, we briefly consider first-passage MC simulations starting from an initial configuration which is periodic and composed of the two motifs `AB' and `AABB', in differing ratios.

The notation `m:n' denotes an initial configuration made by the repetition of the block $(AB)^m(AABB)^n$ to fill the lattice. Fig \ref{crossover} shows the result for studies done for various values of $r=m/n$, on a lattice of size $L=10^7$. One can see a crossover from the behaviour at $r=0$, which is $N(t) \sim t^{-1/2}(\log{(t)})^{-1}$ to the behaviour at $r=\infty$, which is a pure $N(t) \sim t^{-1/2}$ decay. The crossover time seems to grow larger for smaller $r$, but it seems that for long times the behaviour for any $r >0$ will eventually cross over to a $t^{-1/2}$ behaviour, for sufficiently large lattices. The behaviour of more complicated periodic combinations of motifs, and indeed, of random combinations of motifs, is at present unknown. However, based on the above simulations, we expect the existence of two decay regimes even in these cases.

Ódor and Menyhárd \cite{odor} studied the decay for a system which can be mapped to two species annihilation, starting from initial conditions which can be mapped to a random mixtures of AABB and AB blocks, in our notation. As they varied the proportions of different types of blocks in the ic, they saw a change in the decay exponent, from about 0.5 to about 0.55. This is consistent with the crossover phenomenon we observed above, which especially for the relatively small system sizes studied ($L = 24000$) can look like a slowly-varying exponent slightly different from 0.5. Ódor and Menyhárd also argue that this could be an effect of a logarithmic correction to the decay. Lee's results \cite{leeclg14} for the `natural' initial conditions, which show an effective exponent different from and lying between the ones he observed for $n$ even and $n$ odd, can also be explained in this fashion.

\section{Conclusions} We have shown, using the Independent Interval Approximation, and through a mapping to a q-state model, that the effect of exclusion in A-B annihilation in one dimension is to induce kinetic symmetry-breaking in the coarsening process. In particular, we studied initial conditions of the type $A^{n}B^{n}A^{n}B^{n}...$, showing that, surprisingly, the eventual decay depends on the parity of $n$. The number of surviving particles at large times decays as $t^{-1/2}$ for $n$ odd, and as $t^{-1/2}(\log{(t)})^{-1}$ for $n$ even. We also performed first-passage Monte Carlo simulations to verify our claims, and studied the behaviour in a system with a mixture of two types of blocks, showing two time regimes.

These results also explain the results of Lee \cite{leeclg14} on the Conserved Lattice Gas (CLG) in 1D, which maps to the A-B annihilation problem. The CLG consists of particles with an exclusion interaction on a 1D lattice, and the dynamics maps to the A-B annihilation problem (with stationary B particles) with the configurations mapping to each other as $0^{n+1} \rightarrow B^{n}$ and $01^{n+1} \rightarrow \phi A^{n}$, where $\phi$ denotes an empty site. The density $\rho = 0.5$ corresponds to equal densities of A and B particles. $\rho=0.5$ is also the critical density of the model, and hence the decay exponent for two-species annihilation gives the critical decay exponents for activity for the CLG. Based on the results of this paper, one can say that for initial conditions of the form $1^{n+1}0^{n+1}$, one expects a pure power law with exponent $-0.5$ for odd $n$ but a logarithmic correction for even $n$. Lee's measurements \cite{leeclg14} on periodic configurations are consistent with this interpretation. Lee also studied the decay for `natural' initial conditions which can be interpreted to be a mixture of blocks $0011$ and $000111$. A deviation from the expected exponent $0.5$ was also found for this case, lying between the values found for odd $n$ and even $n$, which is consistent with the long-time crossover described in section VI of this paper for mixtures of $AB$ and $AABB$ blocks.

We note that the CLG maps exactly to the A-B annihilation problem with stationary B particles. Thus, the analysis of the critical decay in the CLG within the IIA framework would be the same as ours. However, our numerical results do not carry over exactly, but only indicate that a re-examination of the critical decay in the CLG is necessary. 

These results are also important in light of recent controversies about the effect of initial conditions on critical relaxation in 1D absorbing phase transitions \cite{mohantymanna,kwonkimmanna}, of which the CLG is the simplest solvable example. Kwon and Kim found that a domain structure similar to that of the AB annihilation process can be found in the Manna Sandpile starting from Random Initial Conditions. Whether insights from the AB annihilation process can explain the different exponents found for natural, random and regular initial conditions, remains to be explored.

One can also consider multi-species reactions in 1D \cite{avrahamredner} with periodic initial conditions and same-species exclusion. Perhaps the mapping to a q-state model can be extended to these cases.

\section{Acknowledgments}
I would like to thank Paul Krapivsky for pointing me to very useful references, Kabir Ramola and Francois Landes for helpful discussions and comments on the manuscript, and Deepak Dhar and Kedar Damle for useful discussions. I thank Prof. Géza Ódor for bringing the argument about marginal interactions in his paper \cite{odor} to my attention.


\begin{thebibliography}{11}

\bibitem{zeldovich78} Ya. B. Zeldovich and A. A. Ovchinnikov, Chem. Phys. {\bf 28}, 215 (1978)
\bibitem{toussaintwilzek} D. Toussaint and F. Wilzek, J. Chem. Phys. {\bf 78}, 2642 (1983)
\bibitem{argkop90} P. Argyrakis and R. Kopelman, Phys. Rev. A {\bf 41}, 2121 (1990)
\bibitem{avrahamredner} D. ben-Avraham and S. Redner, Phys. Rev. A {\bf 34}, 501 (1986)
\bibitem{kangredner} K. Kang and S. Redner, Phys. Rev. Lett. {\bf 52}, 955 (1984)\\
K. Kang and S. Redner, Phys. Rev. A {\bf 32}, 435 (1985)
\bibitem{leyvrazredner} F. Leyvraz and S. Redner, Phys. Rev. Lett. {\bf 66}, 2168 (1991)\\
F. Leyvraz and S. Redner, Phys. Rev. A {\bf 46}, 3132 (1992)
\bibitem{singlespec} M.  Bramson  and  D.  Griffeath, Z.  Wahrsch. verw. Gebiete {\bf 53}, 183 (1980)\\
D. C. Torney and H. M. McConnell, J. Phys. Chem. {\bf 87}, 1941 (1983)
\bibitem{derridarevs} B. Derrida, Physics Reports {\bf 301}, 65 (1998)\\
B. Derrida, Pramana {\bf 64}, 695 (2005)
\bibitem{taggedparticle1} T. E. Harris, J. Appl. Probab. {\bf 2}, 323 (1965)
\bibitem{taggedparticle4} R. Arratia, Ann. Probab. {\bf 11}, 362 (1983)
\bibitem{taggedparticle5} P. L. Krapivsky, K. Mallick, and T. Sadhu, Phys. Rev. Lett. {\bf 113}, 078101 (2014)
\bibitem{leeclg14} S. B. Lee, J. Kor. Phys. Soc. {\bf 64}, 1636 (2014)
\bibitem{rossiclg} M. Rossi, R. Pastor-Satorras, and A. Vespignani, Phys. Rev. Lett. {\bf 85}, 1803 (2000)
\bibitem{olivclg} M. J. de Oliveira, Phys. Rev. E {\bf 71}, 016112 (2005)
\bibitem{leelee} S. B. Lee and S.-G. Lee, Phys. Rev. E. {\bf 78}, 040103(R) (2008)\\
S.-G. Lee and S. B. Lee, Complex, Part I, LNICST {\bf 4}, 841 (2009)
\bibitem{jain} K. Jain, Phys. Rev. E {\bf 72}, 017105 (2005)
\bibitem{kwonkimclg} S. Kwon and J. M. Kim, Phys. Rev. E {\bf 90}, 046101 (2014)
\bibitem{odor} G. Ódor and N. Menyhárd, Phys. Rev. E {\bf 61}, 6404 (2000)
\bibitem{majumdardean} S. N. Majumdar, D. S. Dean and P. Grassberger, Phys. Rev. Lett. {\bf 86}, 2301 (2001)\\
S. N. Majumdar and D. S. Dean, Phys. Rev. E {\bf 66}, 056114 (2002)
\bibitem{krapiviia} P. L. Krapivsky and E. Ben-Naim, Phys. Rev. E {\bf 56}, 3788 (1997)
\bibitem{bandyopadhyay} S. Bondyopadhyay, Phys. Rev. E. {\bf 88}, 062125 (2013)
\bibitem{krapivskybook} P. L. Krapivsky, S. Redner, and E. Ben-Naim, A kinetic view of statistical physics (Cambridge University Press, Cambridge) (2010)
\bibitem{pottsderrida} B. Derrida, V. Hakim and V. Pasquier, J. Stat. Phys. {\bf 85}, 763 (1996)
\bibitem{kmc1} T. Opplestrup, V. V. Bulatov, G. H. Gilmer, M. H. Kalos and B. Sadigh, Phys. Rev. Lett. {\bf 97} 230602 (2006)
\bibitem{kmc2} A. Bezzola, B.B. Bales, R.C. Alkire and L.R. Petzold, J. Comput. Phys. {\bf 256}  183 (2014)
\bibitem{mohantymanna} M. Basu, U. Basu, S. Bondyopadhyay, P. K. Mohanty, and H. Hinrichsen, Phys. Rev. Lett. {\bf 109}, 015702 (2012).
\bibitem{kwonkimmanna} S. Kwon and J. M. Kim, Phys. Rev. E {\bf 91}, 012149 (2015)
\bibitem{dandekarthesis} R. Dandekar, PhD Thesis (2014)
\bibitem{kkclgtransition} S. Kwon and J. M. Kim, Phys. Rev. E {\bf 93}, 012106 (2016)
\bibitem{kwonkimmanna2} S. Kwon and J. M. Kim, Phys. Rev. E {\bf 96}, 012146 (2017)

\end{thebibliography}
\end{document}